\documentclass[a4paper,12pt]{article}
\usepackage{listings}
\usepackage[T1]{fontenc}
\usepackage{lipsum}
\usepackage[utf8]{inputenc}
\usepackage{amsmath}
\usepackage[english]{babel}
\usepackage{booktabs} 
\usepackage{multirow}
\usepackage{multicol}
\usepackage{subfigure}
\usepackage{float}
\usepackage{arydshln}
\usepackage{enumitem}
\usepackage{subfiles}
\usepackage{graphicx}
\usepackage{mathtools}
\usepackage{parskip}
\usepackage{array}
\usepackage{multirow}
\usepackage{lineno}
\usepackage{arydshln}
\usepackage{colortbl}
\usepackage{diagbox}
\usepackage{lettrine}
\usepackage{marginnote}
\usepackage{picinpar}
\usepackage{sidecap}
\usepackage{url}

\providecommand{\keywords}[1]
{
  \small	
  \textbf{\textit{Keywords---}} #1
}

\begin{document}

\title{\textbf{A Real Data-Driven Analytical Model to Predict Information Technology Sector Index Price of S\&P 500}}

\author{Jayanta K. Pokharel\thanks{Department of Mathematics and Statistics, University of South Florida, 4202 E Fowler Ave, Tampa, FL 33620, USA.
Email:jpokharel@usf.edu } \and Erasmus Tetteh-Bator\thanks{Department of Mathematics and Statistics, University of South Florida, USA.} \and Chris P. Tsokos \thanks{Department of Mathematics and Statistics, University of South Florida, USA.}
 }
 \date{}
\maketitle

\begin{abstract}
\noindent S\&P 500 Index is one of the most sought after stock indices in the world. In particular, Information Technology Sector of S\&P 500 is the number one business segment of the S\&P 500 in terms of market capital, annual revenue and the number of companies (75) associated with it, and is one of the most attracting areas for many investors due to high percentage annual returns on investment over the years. A non-linear real data-driven analytical model is built to predict the Weekly Closing Price ($WCP$) of the Information Technology Sector Index of S\&P 500 using six financial, four economic indicators and their two way interactions as the attributable entities that drive the stock returns. We rank the statistically significant indicators and their interactions based on the percentage of contribution to the $WCP$ of the Information Technology Sector Index of the S\&P 500 that provides significant information for the beneficiary of the proposed predictive model. The model has the predictive accuracy of 99.4\%, and the paper presents some intriguing findings and the model's usefulness. 
\end{abstract}

.

\section{Introduction}

Capital markets are financial markets in which various long term financial instruments such as long term debt or equity backed securities can be traded \cite{mclindon1996privatization}. Capital markets play vital role to stimulate economic activities of a country. Specifically, capital markets generate two functions. First, it provides opportunity for companies to raise capital from the investors and second, individual investors and firms can maximize their wealth by investing fund in desired financial instruments. The stock market deals only with equity capital, while the capital market deals with equity and debt instruments. Investors are highly attracted to the stock market due to unlimited profit return possibilities. However, high returns are possible at a cost of high risk. Thus, stock investment is called risk-return trade off \cite{stlouisfed}. The fluctuation of stock prices affect the investor’s decision on investing their capital. In the financial markets, stock price movements are governed by the stock price index and it is considered as the references for investors to invest in the capital markets.

The S\&P 500 index is widely considered to be one of the best single gauges for the U.S. equity markets\cite{Global}. It not only reflects the economic activities at the United States but also has greater impact on the global economy. S\&P 500 consists of 11 business sectors which includes Information Technology, Health Care, Financials, Consumer Discretionary, Communication Services, Industrials, Consumer Staples, Energy, Utilities, Real State, and Materials. Information Technology Sector has the largest share (27.9\%) in terms of market capital followed by the Health Care Sector (13\%). Similarly, based on the number of companies listed in the S\&P 500, Information Technology Sector can be ranked number one at the present context as it consists of 75 companies followed by Industrials (73). In the year 2021, it is estimated that the United States tech-sector contributed around 1.8 trillion dollars to the country's GDP, making up approximately 9.3\% of total GDP\cite{kurpayanidi2020problem}.

The Information Technology Sector primarily covers three areas of business and services. Firstly, Technology Software and Services includes companies that primarily develop software in various fields such as the internet, applications, systems, databases management and home entertainment, and companies that provide information technology consulting and services, as well as data processing and outsourced services. Secondly, it covers the field related to technology hardware and equipment, including manufacturers and distributors of communications equipment, computers and peripherals and electronic equipment, and thirdly, semiconductors \& semiconductor equipment manufacturers.

There continuous to be increasing demand of high quality tech-sectors product and services worldwide and hence the sector is significantly growing every year. Prediction of stock index is considered as an important task and is of great attention as predicting stock prices successfully may lead to attractive profits by making proper investment decisions.  It is reported that from 2007 to 2021, the Information Technology Sector recorded the highest aggregate returns of 938.3\% among all 11 business sectors of S\&P 500 followed by the Consumer Discretionary (560.2\%)\cite{Novel}. Therefore, Information Technology Sector Index is one of the most sought after business segments in S\&P 500 for investments.

As more and more individuals and institutions are attracted to the stock markets due to unlimited profit return possibilities, stock price prediction has been of a great deal of interest among academicians, investors and institutions in the equity markets. In this context, we proposed a high quality  real-data driven predictive model to predict the Weekly Closing Price of the Information Technology Sector Index of S\&P 500 with high degree of accuracy. 

We have considered six financial indicators (Beta, Free Cash Flow Per Share, Price-to-Book Ratio, Price Earning Ratio, PEG Ratio \& Dividend Yield)  and four economic indicators (Interest Rate, Index of Consumer Sentiment, US Personal Saving Rate \& US GDP) to build our analytical predictive model which will be described in details in Section 2.1.

The attributable entities (indicators) that are included in the proposed predictive model have a significant relevance in the literature of finance and economics. 'Conditional Relations between Beta and Return', a study by Tang and Shum (2003) found that a measure of stocks volatility of returns which is defined as Beta-factor to have significant effect on returns \cite{tang2003relationships}. However, a study conducted by Sherelene Enriquez-Savery (2016) shows that the Beta-risk factor is wrongly calculated in practice \cite{EnriquezSavery2016StatisticalAO}.  Many researches and business analysts believe that Free Cash Flow Per Share (FCF/Share) and Dividend Yield (Div\textunderscore Yield) are closely associated and play a crucial role in the stock returns. Our study also supports the fact that the Weekly Closing Price of the index is affected by companies' FCF/Share and its interactions with GDP and Dividend Yield. Also, stocks with high dividend yields usually have an advantage of being attractive to many investors due to regular return advantage over their lower-yielding counterparts. Price-to-Book (P/B) Ratio indicates whether the stock is overvalued and can be used in determining the best value of the stock at a given time. We also found that P/B Ratio and its interaction with other economic and financial indicators are statistically significantly contributing factors in the Weekly Closing Price of the index.  Many researchers believe that Price-to-Earning (P/E) Ratio and  Price-Earnings-to-Growth (PEG) Ratio are two of the most crucial indicators that influence the stock returns. Lagevardi \cite{lajevardi2014study} research supports that P/E ratio is more directly related to stock returns than the PEG Ratio and thus, stock returns of the companies is more affected by the P/E ratio as compared to PEG Ratio. On the other hand, many researchers argue that PEG Ratio gives more complete picture of the stock returns  as it also accounts for expected future growth. Our study shows that the PEG Ratio and its interactions with other indicators are statistically more powerful than the P/E Ratio to determine the index price.

As mentioned above, the proposed analytical predictive model also includes four economic indicators. Alam (2009) found that changes of interest rate has a significant negative relationship with changes of the share price \cite{alam2009relationship}. We found that interest rate alone is a weak attributable entity in determining the Weekly Closing Price of the index, however, its interactions with PEG Ratio, P/B Ratio and P/E Ratio have statistically significant effect on the index price. Similarly, many researchers and financial analysts believe that good financial market environment motivates the investors' interest  to keep the growth momentum of the stock market. A study by Lemmon (2006) shows that investors' confidence level (measured by the US index called ICS) exhibits forecasting power for the stock returns \cite{lemmon2006consumer}. Other economic indicators which are used in the developed model are, the US Personal Saving Rate (PSR) and the US GDP.  Studies have shown that current saving rates influences future consumption and supports investments \cite{garner2006should}, thus, it is important to  understand its impact on stocks/index price. Our study also supports the fact that the interaction of PSR with ICS is a statistically significant entity in predicting index price. A study of interlinakge between stock market and GDP growth by Duda, concluded that there is no direct connection between stock market growth and the GDP growth \cite{duda7study}. However, in our study we found that the US GDP is the number one attributable entity in determining the Weekly Closing Price of the Information Technology Sector Index of S\&P 500. In this context, this article presents some intriguing findings and the proposed model's usefulness.

The proposed procedure and methodology can be effectively used to develop predictive models for individual companies and other business sectors of S\&P 500 and we will continue our effort to explore this idea in our future research. 

\section{Methodology}
The detail procedure and the methodology of our proposed predictive model building process is presented in Subsection 2.1 and 2.2 and 2.3. 

\subsection{Data and Description of The Indicator}

Our main goal is to build a data-driven analytical predictive model to predict the Weekly Closing Price of the Information Technology Sector Index of S\&P 500. We collected the required information about our selected indicators using various sources such as Yahoo Finance, FRED Economic Data, Zacks Investment Research, Alpha Query, Morningstar.com, www.wsj.com/market-data, etc. The dataset includes information from January 2017 to December 2019. We collected the weekly information about the indicators for all 75 companies listed in the Information Technology Sector Index of S\&P 500 and averaged the weekly data of each indicator corresponding to these companies to obtain a new weekly data for each attributable indicator in structuring the weekly database of the index. After extensive literature review of the subject area, six financial indicators and four economic indicators are considered that may influence the Weekly Closing Price \textbf{(WCP)} of the index as a measure of our response. 

For the convenience of the reader, we define below the financial indicators from number 1-6, and economic indicators from number 7-10 that are significant entities of our analytical model. 

\begin{enumerate}

\item $\textbf{Beta} (X_{1}):$ The beta value is a statistical measure that compares the volatility of return of a specific stock in relation to those stocks of the market as a whole. In general, stocks with higher beta value are considered to be more riskier, thus, investors will expect higher returns. That is,
\begin{align*}
Beta = \frac{Cov(R_{i},R_{m})}{Var(R_{m})}\ ,
\end{align*}

where $R_{i}$ is the return on individual stock and $R_{m}$ is the return on overall market. $Cov(.,.)$ is the covariance between $R_{i}$ and $R_{m}$ that measures the changes in stock's returns with respect to the changes in market's returns, $Var(.)$ is the variance of the excess market returns over the risk-free rate of returns.\\

\item $\textbf{FCF/Share}(X2):$ Free Cash Flow Per Share (FCF/Share) is a measure of a company’s financial flexibility that is determined by dividing Free Cash Flow  of the company by the total number of shares outstanding. That is,
\[ FCF/Share =
 \frac{\text{Free Cash Flow}}{\text{Number of Share Outstanding}}\ .
\]

\item $\textbf{P/B Ratio} (X_{3}):$ Price-to-Book (P/B) Ratio is a financial ratio used to compare a company's current market value to its book value (book value is the value of all assets minus liabilities owned by a company). That is,

\[ P/B Ratio =
 \frac{\text{Market Value Per Share}}{\text{Book Value Per Share}}\ .
\]

\item  $\textbf{P/E Ratio} (X_{4}):$ Price-to-Earnings (P/E) Ratio is the ratio that measures the current price of a stock concerning its earnings per share and is given by 

 \[ P/E Ratio =
 \frac{\text{Market Value Per Share}}{\text{Earnings Per Share}}\ .
\]

\item   $\textbf{PEG Ratio} (X_{5}):$ Price-Earnings-to-Growth (PEG) Ratio is the stock's Price-to-Earnings (P/E) Ratio divided by the growth rate of its earnings for a specified period and is given by
\[ PEG Ratio =
 \frac{\text{P/E Ratio}}{\text{Annual EPS Growth}}\ .
\]

\item  $\textbf{Dividend Yield}(X6):$ Dividend Yield (Div\textunderscore Yield) is a financial ratio that identifies the percentage of a company's share price that it pays out in dividends each year and is given by

\[ Dividend Yield =
 \frac{\text{Annual Dividend Per Share}}{\text{Current Share Price}}\ .
\]

\item  $\textbf{Interest Rate}(X7):$ The US Federal Fund Rate is used. The term federal funds rate refers to the target interest rate set by the Federal Open Market Committee. This target is the rate at which the Fed suggests commercial banks borrow and lend their excess reserves to each other overnight. \\

\item  $\textbf{US ICS}(X8):$ The Michigan Survey Research Center has developed the Index of Consumer Sentiment (ICS) to measure the confidence or optimisim (passimism) of consumers in their future well-being and coming economic condition. The ICS measures short and long-term expectations of business conditions and the individual's perceived economic well-being. The evidence supports that the ICS is a leading indicator of economic activities as consumer confidence seems to pave the way for major spending and investments.\\

\item  $\textbf{US PSR}(X9):$ The US Bureau of Economic Analysis (BEA) publish the US Personal Saving Rate. The US Personal Saving Rate is the personal saving rate as a percentage of personal income. It is percentage measure of individuals' income left after they pay taxes and expenditures. \\ 

\item  $\textbf{US GDP}(X10):$ Gross Domestic Product (GDP) of the United States (in billion) is used. GDP is the monetary value of all finished goods and services made within a country during a specific period. The components of GDP include personal consumption expenditures (C), business investments (I), government spending (G), exports (X), and imports (M). That is,
\begin{align*}
GDP = C + I + G + (X - M).
\end{align*}
\end{enumerate}

\subsection{Development of Statistical Model}
To develop our multiple regression predictive model, there are several statistical assumptions that our data must satisfy. Then, we proceed to verify these assumptions. 

\noindent {\bf{Normality of the Weekly Closing Price ($WCP$):}} In developing the proposed statistical model for the Information Technology Sector Index Price  of S\&P 500 as a function of the attributable indicators, one of the main assumptions is that the response indicator "$\textbf{WCP}$" should follow the Gaussian Probability Distribution. Normal QQ plot and the normality tests are used to verify the normality of our response, $\textbf{WCP}$.
\begin{figure}[H]
\begin{center}
\includegraphics[width=100mm,scale=0.5]{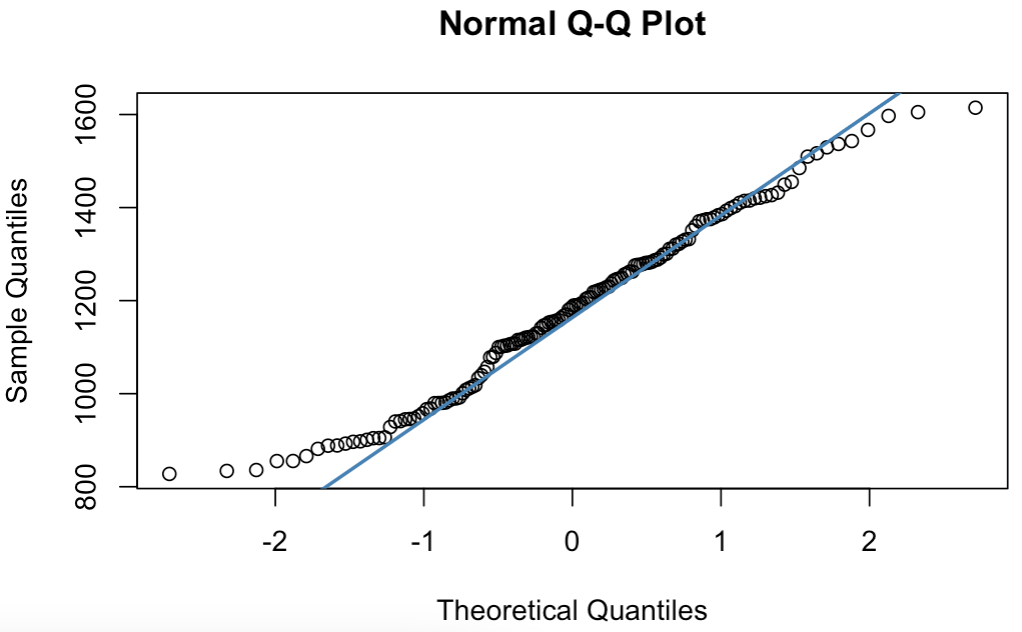}\\[0.1cm]    
\caption{Normal Q-Q Plot of $WCP$}
\label{fig1}
\end{center}
\end{figure}
Figure \ref{fig1} above, shows that there is systematic deviation from normality and the WCP does not entirely follow a Gaussian Probability Distribution. The goodnesss-of-fit test using Shapiro-Wilk normality test yields a p-value = 0.03063 which is less than 0.05 at 5\% level of significance. The normality tests confirm that the response indicator "$\textbf{WCP}$" does not follow the Normal Probability Distribution. Therefore, the Normal Q-Q Plot  of the $WCP$ supports the fact that natural phenomena such as the  Weekly Closing Price of the index does not follow the Gaussian Probability Distribution.
                
In the process of developing a statistical model, our main goal is to express our response $WCP$ in terms of non-linear mathematical function of all significantly contributing indicators including their interactions with high degree of accuracy. Thus, one of the pure forms of the model with all possible interactions and additive error terms that can possibly estimates the $WCP$ of the index is expressed as follows:
\begin{equation} \label{eq1}
\begin{split}
WCP = \alpha + \sum\limits_{i}\beta_{i}x_{i} + \sum\limits_{i\neq j}\rho_{ij}x_{i}x_{j} + \epsilon
\end{split}
\end{equation}
where,
\begin{itemize}
 \item $\alpha$ is the intercept of the regression model,
 \item $\beta_{i}$ is the  coefficient of $i^{th}$ individual indicator $x_{i}$,
 \item $\rho_{ij}$ is the coefficient of $ij^{th}$ interaction term $x_{i}x_{j}$,
 \item $\epsilon\overset{\mathrm{iid}}{\sim} N(0,\sigma^2)$ is a Gaussian error terms (residual error).
\end{itemize}

The main assumption behind constructing the above model is that the response indicator "$\textbf{WCP}$" should follow the Gaussian Probability Distribution. However, we noticed that $WCP$ does not support it. Therefore, we must apply a non-linear transformation to WCP and see if the transformation can adjust the scale of the response to follow the Normal Probability Distribution. We used  the Johnson Transformation bounded family\cite{chou1998transforming} to transform $WCP$. Given a continuous random variable X whose distribution is unknown and is to be approximated, Johnson proposed three normalizing transformations that has the following general form: 
\begin{equation}\label{JT}
Z = \gamma + \eta ln \left[  \frac{ (X - \xi) } { ( \lambda + \xi - X)}  \right],\ \xi< X < \xi + \lambda
\end{equation}
where,
\begin{center}
$-\infty<\gamma<\infty$,\ $\gamma$ - the shape parameter,\\
$\eta > 0$,\ $\eta$ - the shape parameter,\\ 
$\lambda > 0$,\ $\lambda$ - the scale parameter, \\
$-\infty<\xi<\infty$,\ $\xi$ - the location parameter.
\end{center}

After we applied Johnson Transformation (\ref{JT}) on $WCP$, we obtained Equation (\ref{JTWC}) to estimate the new Transformed Response Indicator ($WCP^T$). 

\begin{equation}\label{JTWC}
WCP^T = 0.4091 + 1.2208  \ ln \left[  \frac{ (X - 711.5838) } { ( 1794.5468 - X)}  \right],
\end{equation} 

where $\gamma$ = 0.4091, $\eta$ = 1.2208, $\xi$ = 711.5838 and $\lambda$ = 1082.963.\\

\begin{figure}[H]
\begin{center}
\includegraphics[width=100mm,scale=0.5]{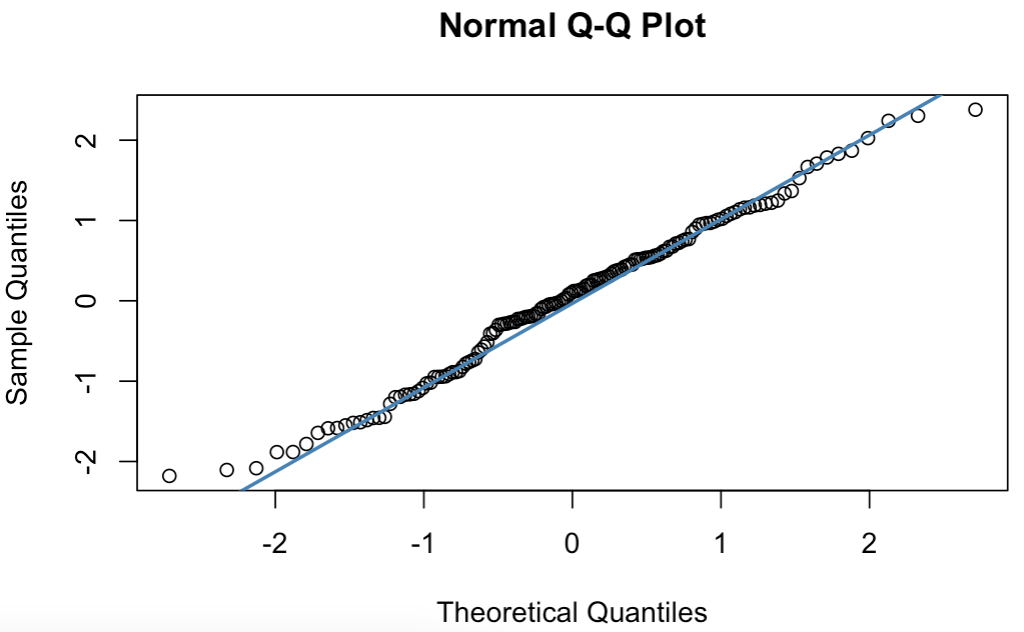}\\[0.1cm] \vspace{0mm} 
\end{center}
\caption{Normal Q-Q Plot of $WCP^T$}
\label{fig2}
\end{figure}
We then checked the normality of the $WCP^T$ with the help of Normal Q-Q Plot and normality tests. The Figure \ref{fig2} above, shows that there is no systematic deviation from normality and, thus, supports the fact that it follows the Gaussian Probability Distribution. The goodnesss-of-fit test using Shapiro-Wilk normality test yields the p-value = 0.3031 which is much greater than 0.05 at 5\% level of significance. Also, the Anderson-Darling normality test has the p-value = 0.271 which is greater than 0.05. Both of these normality tests confirm that the new transformed response indicator ${WCP^T}$ does  follow the Normal Probability Distribution.

Here onward, we use the transformed response indicator "$WCP^T$" as the new response indicator to conduct our statistical analysis and later, we must apply anti-Johnson transformation to obtain the original scale of $WCP$. 

\noindent {\bf{Non-Linear Relationship Among Indicators:}} Multiple linear regression assumes that there is little or no multicollinearity in the data and correlation analysis is an important part of the model building process. Correlation analysis contributes to the understanding of economic behavior by locating the critically important attributable indicators on which other indicators depend. There should be non-linear relationship among our indicators used in the model. The Figure \ref{fig3} below, shows the strength of linearity among all possible attributable indicators. 
\begin{figure}[H]
\begin{center} 
\includegraphics[width=110mm,scale=0.5]{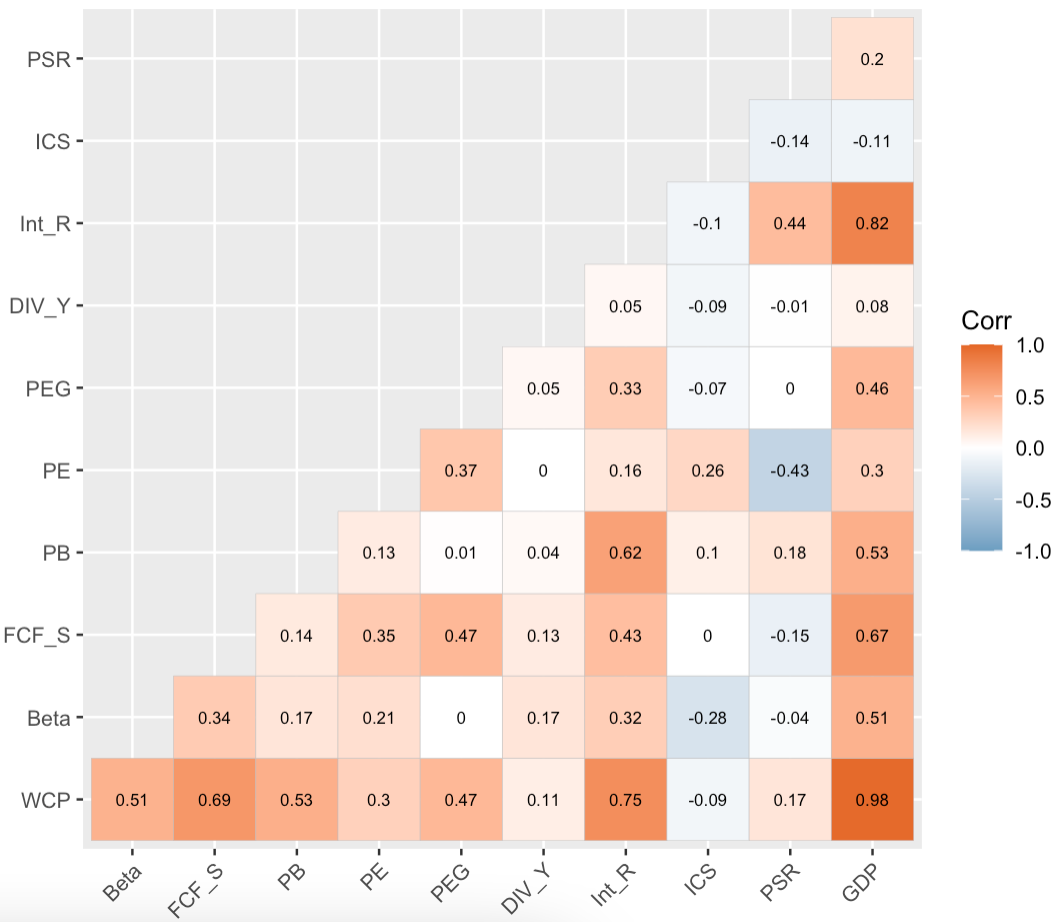}\\[0.1cm] 
\caption{Correlation Matrix of All Possible Attributable Indicators}
\label{fig3}
\end{center}
\end{figure}
We observed that most of our attributable indicators are linearly associated with $\textbf{WCP}$ which is a good sign to conduct a multiple regression for the given dataset. The correlation coefficient between the GDP and the Interest Rate is 0.82 which is the matter of little concerned, where as the correlation coefficient  among other attributable indicators are within the acceptable range at the cut off value of 0.80.  

We then checked the Variance Inflation Factor (VIF) by constructing a multiple regression model to statistically verify the existence of multicollinearity among our attributable indicators. VIF is a measure of the amount of multicollinearity present in a set of multiple regression variables. The VIF score less than 10 indicates that the dataset has no multicollinearity effect among the attributable indicators, otherwise they are linearly correlated.
\begin{table}[H]
\begin{center}
\begin{tabular}{|c|c|c|c|c|c|c|c|c|c|}
\hline
$X_{1}$ &  $X_{2}$  &  $X_{3}$ &  $X_{4}$ &  $X_{5}$ &  $X_{6}$ &  $X_{7}$ &  $X_{8}$ &  $X_{9}$ &  $X_{10}$ \\
\hline
1.93 & 2.58 & 2.02 & 1.80 & 1.84 & 1.06 & 4.39 & 1.38 & 1.91 & 6.92 \\
\hline
\end{tabular}
\label{t1}
\caption{Variance Inflation Factor (VIF)}
\end{center}
\end{table}

We observed from the Table 1 above, that VIF value for each attributable indicators are far less than 10 and thus, our dataset passed the multicollinearity test. The outcome of the VIF satisfies the nonlinearity assumption of multiple regression for our dataset.

Now, we proceed to develop the predictive model using ordinary least squared methods and details are presented in Subsection 2.3.

\subsection{Fitting the Statistical Model}
  
During our data preprocessing step, we observe that our attributable indicators are in different scale and range. Therefore, feature scaling (the process of normalizing the range of features/column vectors of different units in a dataset) is used to standardize the range of our attributable indicators. We randomly split the data into train and test dataset in the ratio of 80:20. As a statistical model building process on the train dataset, we proceeded to estimate the coefficients (weights) of the actual contributable indicators for the transformed data in Equation (\ref{JTWC}). We began with the ordinary least square methods with 10 indicators as described in subsection 2.1. We then run the full statistical model including all possible two way interaction terms (${n \choose x}$ = 45, where n = 10 and k = 2). Since backward elimination is deemed one of the best traditional methods for a small set of feature vectors to handle the problem of overfitting, we then proceeded to determine the most significant individual indicators and interactions  using stepwise backward elimination method. After careful considerations, we observed five attributable indicators and nine interaction terms remained statistically significant in our final model.  The P/B Ratio, PEG Ratio and GDP are found to be highly significant and, FCF/Share and P/E Ratio are moderately significant at 5\% level of significance. Similarly, the interactions between Beta and P/B Ratio, FCF/Share and GDP, P/B Ratio and Interest Rate, PEG Ratio and Interest Rate, and PEG Ratio and GDP are highly significant where as  Beta and PSR, FCF/Share and Dividend Yield, P/E Ratio and Interest Rate, and, ICS and PSR are moderately significant at 5\% level of significance. We noticed that some of those indicators that are not significant individually, their interactions with other indicators are highly or moderately significant and thus, we decided to keep them in the proposed model. The trained model has the R-squared of 0.9944 and the adj. R-squared of 0.9934. The high R-Squared value indicates that the proposed model is of high quality model with excellent predictive accuracy of 99.4\%.

We further evaluated the predictive model based on multiple regression assumptions, performance accuracy using various metrics, and validated using k-Fold Cross Validation, and detail discussions are presented in Section 3.

The best preferred statistical model with all significantly contributable indicators and their interactions that estimates the Transformed Weekly Closing Price ($\widehat{WCP^T}$) of the Information Technology Index of S\&P 500 is given by Equation (\ref{trPrice}).

\begin{equation}\label{trPrice}
\begin{split}
\widehat{WCP^T}\ =  \ & 0.0732 + 0.0070\ X_{1} - 0.0504\ X_{2} + 0.1067\ X_{3} + 0.0578\ X_{4} + 0.0941\ X_{5} - 0.0041\ X_{6}\\
& - 0.0168\ X_{7} - 0.0120\ X_{8} + 0.0211\ X_{9}
+ 0.6781\ X_{10} + 0.1013\ X_{1} X_{3} -  0.0864\ X_{1} X_{9} \\
& + 0.0351\ X_{2} X_{6} - 0.0989\ X_{2}X_{10}
- 0.1139\ X_{2} X_{7} - 0.0380\ X_{4} X_{7} - 0.6059\ X_{5} X_{7} \\
& + 0.6054\ X_{5} X_{10} + 0.0331\ X_{8} X_{9},
\end{split}
\end{equation}
where,\\

$\widehat{WCP^T}$ \ = Transformed Average Weekly Closing Price, \\
$X_{1} = Beta$,\ $X_{2} = FCF/Share$,\ $X_{3} = P/B\ Ratio$,\ $X_{4} = P/E\ Ratio$, $X_{5} = PEG\ Ratio$,\\
$X_{6}$ = Div\textunderscore Yield, \ $X_{7}$ = Int\textunderscore Rate,\ $X_{8} = ICS$,\ $X_{9} = PSR$,\ $X_{10} = GDP$.

Equation (\ref{trPrice}) estimates the Transformed Weekly Closing Price $(WCP^T)$ of the index based on Johnson transformation. Here, anti-transformation on Equation (\ref{JTWC}) is required to estimate the desired prediction of $WCP$ of the index, and it is given by the following Equation (\ref{anjohn}) or (\ref{antijohn}).

\begin{equation}\label{anjohn}
\widehat{WCP}\ =  \xi + \frac{\lambda}{1+ exp\left( \frac{\gamma - \widehat{WCP^T}}{\eta} \right)}
\end{equation}
Or,
\begin{equation}\label{antijohn}
\widehat{WCP}\ =  \ 711.5838 + \frac{1082.963}{1+ exp\left( \frac{0.4091 - \widehat{WCP^T}}{1.2208} \right)}.
\end{equation}

Equation (\ref{antijohn}) represents the best preferred statistical predictive model to estimate the Weekly Closing Price ($WCP$) of the Information Technology Sector Index of S\&P 500. Given any standardized values of the attributable indicators, the Equation (\ref{trPrice}) estimates the $\widehat{WCP^T}$ and then we substitute the value of $\widehat{WCP^T}$ in the analytical model represented by Equation (\ref{antijohn}) to predict the Weekly Closing Price of the Information Technology Sector Index of S\&P 500.

\section{ Evaluation of the Predictive Model:} In this section we proceed to evaluate the quality of the proposed model.

\noindent {\emph{Mean Residuals:}} The difference between the observed value "$y$" of the attributable indicator and the predicted value "$\hat{y}$" is called the residual ($\hat{e}$). 
\begin{equation}
Residual\ (\hat{e}) = Observed \ value \ (WCP) - \ Predicted \ value \ (\widehat{WCP}).
\end{equation}

Both the sum and the mean of the residuals should be equal to zero assuming that the regression line is actually the line of "best fit". It is found that the sum of the residuals and the mean residuals values are $e^{-16} \approx 0$ \ and $e^{-18} \approx 0$, respectively. Thus, it satisfies the assumption that the regression line is actually the line of "best fit".

\noindent {\emph{Normality of Residuals:}} One of the key assumptions to verify of the proposed model is the normality of the residuals. The following Figure 4(a), QQ-Plot of the Residuals shows that residuals are closely fit the regression line and Figure 4(b) the Distribution of the Studentized Residuals depicts that the residuals are approximately normally distributed. The p-value from normality test using Shapiro-Wilk and Anderson-Darling tests are 0.2091 and 0.1482, respectively. Both p-values are greater than 0.05 which indicate that normality assumptions of the residuals are satisfied at 5\% level of significance.\\ 

\begin{figure}[H]
\begin{center}  
\includegraphics[width=130mm,scale=0.5]{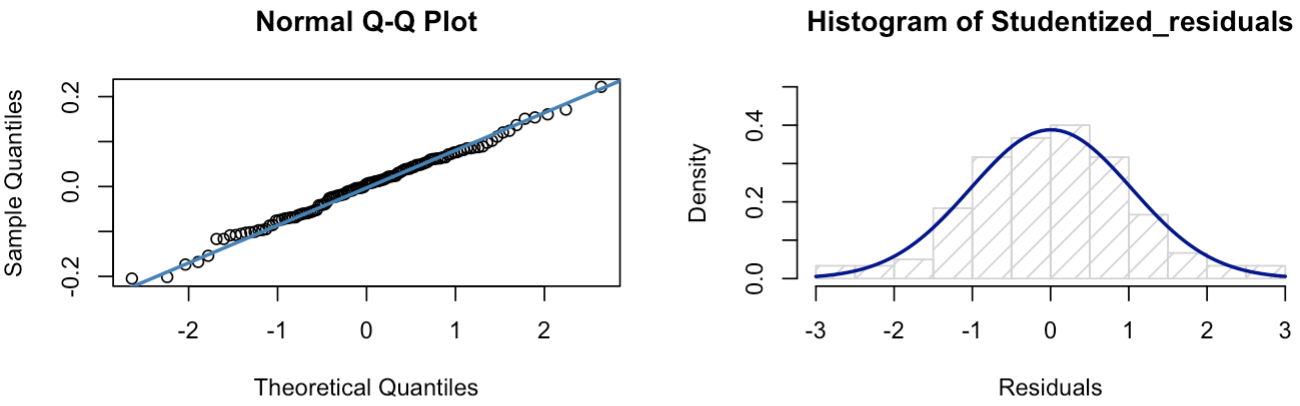}\\[0.1cm] 
\caption{ 4(a): Q-Q Plot of Residuals\ 4(b): Distribution of Stud\textunderscore Residuals}
\label{fig4}
\end{center}
\end{figure}

{\emph{Homoscedasticity:}} One of the main assumptions for the regression model is the homogeneity of the variance of the residuals. Figure \ref{fig5} below, shows that there is no definite pattern of the residuals and they are approximately equally distributed on either sides of reference line. This indicates that the proposed statistical model satisfies the assumption of constant variance of the residuals.

\begin{figure}[H]
\begin{center}
\includegraphics[width=110mm,scale=0.5]{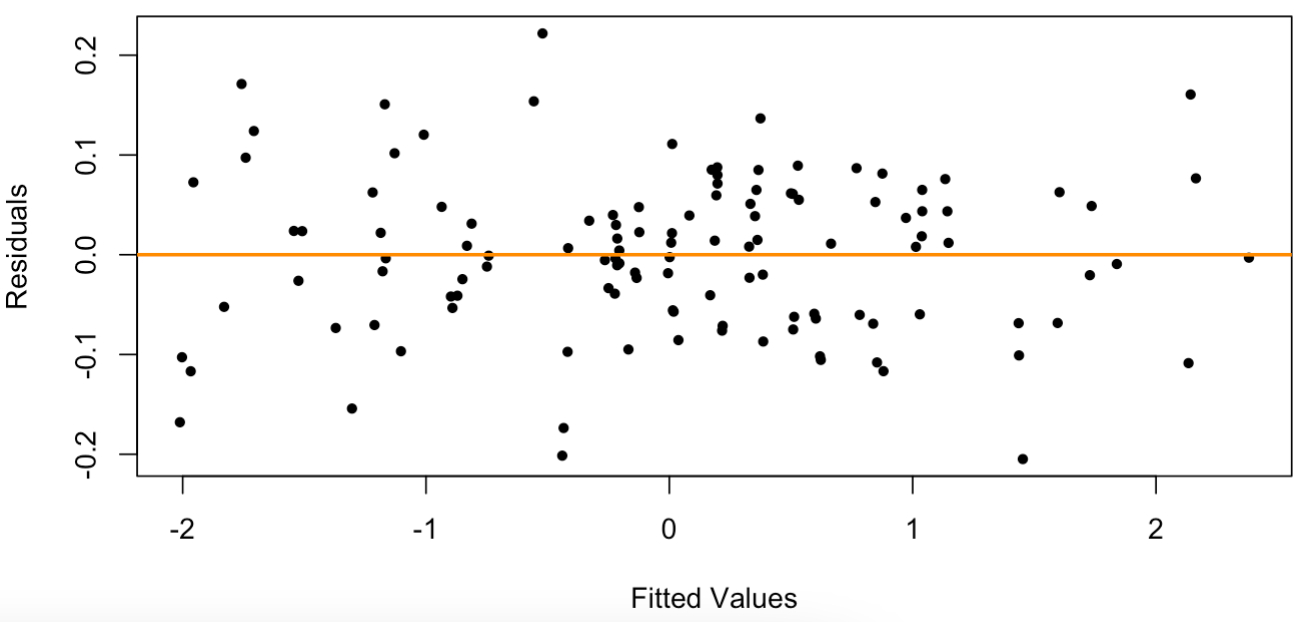}\\[0.1cm] 
\caption{Residuals Vs. Fitted Values}
\label{fig5}
\end{center}
\end{figure}

{\emph{Model Performance}:} The proposed model is evaluated based on various performance metrics.

Firstly, the proposed non-linear predictive model is evaluated based on the coefficient of determination $R^2$, and the adjusted $R^2$, which are important criteria to judge model accuracy. The $R^2$ is estimated by
\begin{equation}
R^2\ = 1 - \frac{SSE}{SST},
\end{equation}
where $SST = SSR + SSE$. The regression sum of squares ($SSR$) that measures the variation on the response explained by the model and the sum of squared errors ($SSE$), also called as the residual sum of squares, measures the variation remain unexplained by the model. The proposed predictive model has the R-squared of 99.4\%. 

In addition, the adjusted R-Squared is given by 
\begin{equation}
R^2_{adj}\ = 1 - \left [\frac{(1 - R^2) (n-1)}{n-k-1}\right],
\end{equation}
where n is the number of data points and k is the number of predictors in the model. The adjusted R-squared of the proposed model is 99.3\% which is pretty close to R-squared. The closer the R-squared and the adjusted R-squared value are, the more accurate the model is, that insures the predictive accuracy of the proposed model.

Secondly, the root mean square error (RMSE) is one of the most widely used measures to evaluate model performance. RMSE is given by\\
\begin{equation}
RMSE \ = \sqrt{\frac{\sum \limits_{i} ({WCP_{i}} - \widehat{WCP_{i}})^2}{N}},
\end{equation}
where N is the number of data points, ${WCP_{i}}$ is the $i^{th}$ observation, and $\widehat{WCP_{i}}$ is its corresponding prediction. The RMSE of the proposed model on our test dataset is 22.72.

The RMSE for the train and test model are found to be  0.0748 and 0.0816, respectively. The difference in RMSE between train and test dataset is 0.0068 (very small). This results ensures the goodness-of-fit of our  high quality model for unseen data for accurate prediction. \\

Thirdly, mean absolute percentage error (MAPE) is another useful metric for the model evaluation. MAPE is calculated by
\begin{equation}
MAPE \ = \left(\frac{1}{N}\sum \limits_{i} \bigg | \frac{WCP_{i} -\widehat{WCP_{i}}) }{WCP_{i}}\bigg |\right)\times100.
\end{equation}
The mean absolute percentage error (MAPE) based on observed and predicted values is 1.38\%. Given the nature of the data, this is relatively small value and it ensures the quality of the proposed model.

We also used relative root mean square error (RRMSE) to evaluate the model performance. RRMSE is a dimensionless form of RMSE. RRMSE is the root mean squared error normalized by the root mean square value where each residual is scaled against the actual value. While RMSE is restricted by the scale of the original measurements, RRMSE can be used to compare different measurement techniques. RRMSE is given by
\begin{equation}
RRMSE \ = \sqrt{\frac{\frac{1}{N}\sum \limits_{i} (WCP_{i} - \widehat{WCP_{i}})^2}{\sum \limits_{i} (\widehat{WCP}_{i})^2}}.
\end{equation}

In general, RRMSE less than 10\% is considered to be excellent model performance. The proposed model has the RRMSE of 1.89\% which very low by standard and it ensures the predictive capability of the model with a high degree of accuracy, that is 99.4\%.

\noindent {\emph{k-Fold Cross Validation}:} We performed k-fold repeated cross validation on training dataset to evaluate the goodness-of-fit of the proposed model such that the test error gives an idea about the predictive consistency of the analytical model on a new dataset. We used k = 10, that is 10-fold cross-validation, a re-sampling technique that randomly divides the training data into 10 groups/folds of approximately equal size. The model is fit on 9 (i.e. k-1) folds and then the remaining fold is used to compute model performance. This procedure is repeated 10 times; each time, a different fold is treated as the validation set. If the current selected model has a good predicted power, then the mean square error for the training data should be approximately equal to the mean square predictive error of the test data.

In our proposed analytical model, the Mean Squared Error for training data (MSETr) and Mean Squared Predictive Error on the test data (MSPE) are 0.0065554 and 0.0061372 respectively. The MSETr and MSPE are approximately equal, therefore, we are confident that the proposed analytical model does not suffer from the issues of overfit  or underfit, rather it is the best fit predictive model. 

\section{Usefulness of the Proposed Highly Accurate Predictive Model}

The proposed highly accurate predictive model which is represented by Equation (\ref{antijohn}) in Subsection 2.3 has its own significant importance in the field of financial research and applied statistics. 
In this section, we discuss the usefulness of the proposed analytical model. Here, we illustrates the five important usefulness of the proposed model. 
\begin{enumerate}
    \item It identifies the most statistically significant indicators that drive the $WCP$ of the Information Technology Sector Index of S\&P 500. In this study FCF/Share, P/B Ratio, P/E Ratio, PEG Ratio and the US GDP are the most influential individual indicators to predict $WCP$ of the index.

    \item It also identifies the important interactions of the indicators that significantly contribute to the $WCP$ of the index. We observe that the interaction between Beta and P/B Ratio, Beta and PSR, FCF/Share and Dividend Yield, FCF/Share and GDP, P/B Ratio and Interest Rate, PE Ratio and Interest Rate, PEG Ratio and Interest Rate, PEG Ratio and GDP and, ICS and PSR significantly effect the estimates of the $WCP$ of the index. It is interesting to note that PEG Ratio and its interactions with other indicators such as Interest Rate, GDP and PSR, are the most influential contributing factors to predict the $WCP$ of the index. 
    
    \item It identifies the rankings of the indicators and their interactions based on the percentage of contribution to the $WCP$ of the index as shown in Table \ref{t2}, below.
    
\begin{table}[H]
\begin{center}
\begin{tabular}{|c|c|c|}
\hline
\textbf{Rank} & \textbf{Indicators/Interaction}  & \textbf{Contributions\ \%} \\
\hline
1 & GDP & 18.60 \\ 
2 & PEG Ratio \ $\cap$ \ Int\textunderscore R & 14.23\\
3 & PEG Ratio\ $\cap$ \ GDP & 11.65\\
4 & P/B Ratio & 7.32 \\
5 & PEG Ratio & 7.06 \\
6 & Beta\ $\cap$ \ P/B \ Ratio & 7.00 \\ 
7 & P/B Ratio\ $\cap$ \ Int\textunderscore R & 6.42 \\
8 & FCF/Share \ $\cap$ \ GDP & 4.29 \\
9 & ICS \ $\cap$ \ PSR & 3.93 \\
10 & Beta \ $\cap$ \ PSR & 3.56 \\
11 & P/E Ratio & 3.40 \\
12 & P/E Ratio\ $\cap$ \ Int\textunderscore R & 2.90 \\
13 & FCF/Share & 2.89 \\
14 & FCF/Share\ $\cap$ \ DIV\textunderscore Y & 2.89 \\
15 & PSR & 1.71 \\
16 & ICS & 1.12 \\
17 & Div\textunderscore Y & 0.54 \\
18 & Int\textunderscore R & 0.49 \\
19 & Beta & 0.37 \\
\hline
\end{tabular}
\caption{Rank of the Most Significant Indicators and their Interactions based on the Percentage of Contribution to the $WCP$ of the Index}
\label{t2}
\end{center}
\end{table}

    We see that GDP contributes the highest percentage 18.60\% to the $WCP$ of the index followed by the interaction between PEG Ratio and Interest Rate, 14.23\%. Our findings from this study about the GDP, as the most influential indicator to determine $WCP$ of the index does not support the argument made by Duda\cite{duda7study}. He reported that there is no direct connection between stock market growth and the GDP growth. However, we found that the US GDP is the number one contributing attributable indicators among the above mentioned economic and financial indicators, and it is highly positively correlated with $WCP$ of the index, thus, GDP growth directly impacts stock market growth. It makes sense with the fact that growing GDP implies growth in per capita income and thus, investors will have more money to spend/invest. This enhances economic activities and stimulates stock market as well. Similarly, it is interesting observation to note that the Dividend Yield is one of the the least contributing indicators (0.54\%) among the list of attributable indicators included in this study to predict the $WCP$ of the index. However, its interaction with FCF/Share is significant entity of the proposed model. These finding makes sense with the fact that most of the tech-company have not distributed dividend in the past and investors solely do not rely on dividend only to invest on tech-sector's stocks but they also look for growth in their investment.
    
    In addition the PEG Ratio is found to be more powerful indicator than the P/E Ratio to predict the $WCP$ of the index. The interaction of the PEG Ratio with Interest Rate and the US GDP are two major contributing factors to estimate the index price. This findings also contradicts the results presented by Lagevardi \cite{lajevardi2014study}. The difference in the findings may be the outcome of the comparison of highly different stock indices. We notice that PEG Ratio is more directly related to the $WCP$ of the Information Technology Sector Index of S\&P 500 than the P/E Ratio and thus, stock returns of the Information Technology Sector Index of S\&P 500 is more affected by the PEG Ratio as compared to the P/E Ratio.
    
    The ranking of indicators included in the model based on the percentage of contribution to the response is extremely important as it provides significant information to the beneficiary of the predictive model. Having prior knowledge of the strength of attributable indicators and their interactions that significantly contribute to the response is extremely important while conducting any survey in the subject area or making important financial decisions. The rank of the indicators based on the percentage contribution of the attributable indicators and their interactions to the $WCP$ of the index serves as a prior knowledge for scientific researchers, business analysts and investors. It also updates intuitive knowledge in the mind of consumers and service providers so that they can focus more on those influential indicators while making important business decisions.
    
    \item For any given set of indicators, the proposed model predicts the change on $WCP$  accurately. 
\begin{table}[H]
\begin{center}
\newcolumntype{C}[1]{>{\centering\arraybackslash}m{#1}}
\begin{tabular}{ |C{1cm}|C{1.9cm}|C{1.9cm}|}
\hline
\textbf{Obs.} & \textbf{Observed}  & \textbf{Predicted} \\
\hline
6 & 865.88 & 862.88 \\
12 & 897.57 & 892.78 \\
15 & 904.64 & 927.23 \\
23 & 951.57 & 949.24 \\
24 & 957.79 & 944.71 \\
27 & 979.69 & 976.07  \\
28 & 979.93 & 982.68 \\
39 & 1018.15 & 1019.11 \\
42 & 1047.08 & 1066.12 \\
46 & 1087.60 & 1046.87 \\
49 & 1103.12 & 1110.35 \\
65 & 1157.09 & 1131.20 \\
69 & 1180.28 & 1161.03 \\
74 & 1184.62 & 1178.35 \\
75 & 1203.88 & 1187.35\\
\hline
\end{tabular}
\begin{tabular}{ |C{1cm}|C{1.8cm}|C{1.8cm}|}
\hline
\textbf{Obs.} & \textbf{Observed}  & \textbf{Predicted} \\
\hline
81 & 1206.58 & 1211.22 \\
82 & 1279.16 & 1211.22 \\
103 & 1283.32 & 1279.86 \\
106 & 1300.75 & 1293.12 \\
111 & 1311.17 & 1330.36 \\
112 & 1320.19 & 1315.09 \\
114 & 1370.62 & 1339.02 \\
121 & 1374.82 & 1351.95 \\
123 & 1378.90 & 1358.36 \\
125 & 1410.34 & 1387.87 \\
131 & 1415.14 & 1396.16 \\
133 & 1421.26 & 1416.56 \\
135 & 1424.52 & 1412.52 \\
137 & 1426.76 & 1456.27 \\
146 & 1542.98 & 1530.36\\
\hline
\end{tabular}
\caption{The List of Observed and Predicted Values of the $AWCP$}
\label{t3}
\end{center}
\end{table}
Table \ref{t3} above, shows the list of Observed Vs. Predicted values of the $WCP$. We can clearly see that the predicted values are very close to the observed values and, thus attests to the accuracy of our proposed model's predictive power of 99.4\ \%. The proposed analytical model can be helpful for the researchers, economists and financial analysts to understand how the $WCP$ of the index varies when any one of the attributable indicators is varied, keeping the other indicators fixed. In other words, understanding the behaviour of the attributable indicators and their interactions help predict the change in the Weekly Closing Price of the Information Technology Index of S\&P 500.

\item Having such an excellent model, the proposed procedure and methodology can be effectively used to develop predictive models for individual companies and other business sectors of the S\&P 500 to predict their closing price, thereby facilitating companies, business analysts and investors to make effective financial decisions.
\end{enumerate}

\section{ Conclusion}

The proposed real data-driven analytical model has its own significance in the field of applied finance and economics. It is developed using strong theoretical understanding of the statistical concepts and financial domain knowledge. The predictive model satisfies all the statistical assumptions, has been tested and validated and, ensures the predictive accuracy of 99.4\ \%. The highlights of the usefulness of the model presented in the Section 4 is the testimony of the quality of the proposed model and its unique contribution to the field of applied finance and economics.

The paper presents some intriguing findings about the attributable indicators and their interactions that influence the Weekly Closing Price of the index. For instance, individual indicator, the US GDP is the most powerful economic indicators (18.60\ \%) followed by P/B Ratio\ (7.32\ \%) and PEG Ratio\ (7.06\ \%) based on the percentage of contribution to the Weekly Closing Price of the index.  Similarly, PEG Ratio and its interactions with the Interest Rate and the US GDP contribute closely 33\ \% to the $WCP$ of the index. Much research has been conducted in the past to understand the impact of financial and economic indicators on stock returns, however the study of the interactions effect of those indicators on the stock/index price is hardly been explored. Our proposed analytical model not only identifies the most contributing attributable indicators and their interactions but also recognizes the relative importance of those indicators while predicting the Weekly Closing Price of the Information Technology Sector Index of S\&P 500.

As more and more individual investors and institutions are attracted to the stock market due to unlimited profit possibilities, stock price prediction has been of a great deal of interest among academician, investors and institutions and, our proposed high quality predictive model serves this purpose. The proposed model is very useful for individual investors and institutions to assess the short and long-term investment strategies. It is also equally important for the companies' managers and shareholders to build policies and strategies to keep up the momentum of the index price by closely monitoring the key indicators/interactions of their individual stock that directly affects the performance of the Information Technology Sector Index of S\&P 500.

The proposed model building procedure and methodology can be effectively used to develop predictive models for individual companies and other business sectors of S\&P 500 and we will continue our effort to explore this idea in our future research.\\

\section{Statements and Declarations}

To the best of our knowledge, this manuscript represents our original work and meets the ethical standards set by the Committee on Publication Ethics (COPE).

\textbf{Ethical approval:} 
\begin{itemize}
\item We all authors, Jayanta K. Pokharel, Erasmus Tetteh-Bator and Chris P. Tsokos have agreed for authorship, read and approved the manuscript, and given consent for submission and subsequent publication of the manuscript. Mr. Pokharel acts as the corresponding author for this manuscript.
\item We ensure that the manuscript and its contents are not a duplicate publication and it is completely original.
\item We ensure that the manuscript represents our original work and meets the ethical standards set by the Committee on Publication Ethics (COPE).
\end{itemize}
\textbf{Acknowledgement:}
Not applicable

\textbf{Consent for publication:} Authors give full consent to SN Business \& Economics to publish this manuscript.

\textbf{Author contribution:}
We declared that Mr. Jayanta K. Pokharel is the main author of this manuscript who designed the concept, prepared dataset, analyzed data and wrote the manuscript. Mr. Erasmus Tetteh-Bator supported the main author while preparing dataset using various web sources, data cleaning and proof reading of the manuscript. Dr. Chris P. Tsokos, Distinguished University Professor of Mathematics \& Statistics at USF is the major professor of Mr. Pokharel's doctoral dissertation committee and Dr. Tsokos guided the authors by approving the concept design, proposed procedures and methodologies and, monitoring the quality of work. 

\textbf{Funding:}
Not applicable

\textbf{Conflicts of Interest:} The authors declare that there are no conflicts of interest regarding the publication of this article.

\textbf{Data availability statement}: The required dataset is prepared using various sources such as Yahoo Finance, FRED Economic Data, Zacks Investment Research, Alpha Query, Morningstar.com, www.wsj.com/market-data, etc. Authors are ready to submit the dataset to SN Business \& Economics as needed.

\textbf{Disclosure}: This research is protected by the University of South Florida TTO.


\bibliographystyle{unsrt}
\bibliography{reference}
\end{document}